\documentstyle[12pt,epsfig]{article} 
\newcommand{\beq}{\begin{equation}} 
\newcommand{\eeq}{\end{equation}} 
\newcommand{\beqa}{\begin{eqnarray}} 
\newcommand{\eeqa}{\end{eqnarray}} 
\newcommand{\beqan}{\begin{eqnarray*}} 
\newcommand{\eeqan}{\end{eqnarray*}} 
\newcommand{\ba}{\begin{array}} 
\newcommand{\ea}{\end{array}} 
\newcommand{\no}{\nonumber}

\newcommand{\ve}{\varepsilon}

\newcommand{\co}{\; \; ,} 
\newcommand{\nn}{\nonumber \\} 
\newcommand{\scs}{\co \;}

\newcommand{\bea}{\begin{eqnarray}} 
\newcommand{\eea}{\end{eqnarray}}

\newcommand{\epe}{\ve_{\pi^0\eta}} 
\newcommand{\pe}{$\pi^0$-$\eta$} 
 
\newcommand{\epoe}{\varepsilon'/\varepsilon} 
\newcommand{\RE}{\mbox{\rm Re}} 
\newcommand{\IM}{\mbox{\rm Im}} 
\newcommand{\hepph}[1]{{\tt hep-ph/#1}} 
\newcommand{\hepex}[1]{{\tt hep-ex/#1}} 
\newcommand{\PL}[3]{{Phys. Lett.} {#1} {(19#2)} {#3}}

\newcommand{\NP}[3]{{Nucl. Phys.} {#1} {(19#2)} {#3}} 
\normalsize 
\begin{document} 
\begin{titlepage} 
\begin{flushright} 
UWThPh-1999-74\\ 
FTUV/99-82\\ 
IFIC/99-86\\ 
Dec. 1999\\ 
\end{flushright} 
\vspace{2.5cm} 
\begin{center} 
{\Large \bf \pe \, Mixing and CP Violation*}\\[40pt] 
G. Ecker$^{1}$, G. M\"uller$^{1}$, H. Neufeld$^{1,2}$  
and A. Pich$^{2}$ 
 
\vspace{1cm} 
${}^{1)}$ Institut f\"ur Theoretische Physik, Universit\"at 
Wien\\ Boltzmanngasse 5, A-1090 Vienna, Austria \\[10pt] 
 
${}^{2)}$ Departament de F\'{\i}sica Te\`orica, IFIC, Universitat de 
Val\`encia - CSIC\\ 
Apt. Correus 2085, E-46071 Val\`encia, Spain 
 
\vfill 
{\bf Abstract} \\ 
\end{center} 
\noindent 
We discuss \pe \, mixing and its implication for 
$\ve'/\ve$ to next-to-leading order in the low-energy expansion.  
The big effect due to $\eta$-$\eta'$ mixing is shown to be largely 
cancelled by other contributions occurring at the same order 
in the chiral expansion.  
 
\vfill 
\noindent * Work supported in part by TMR, EC-Contract  
No. ERBFMRX-CT980169 (EURODA$\Phi$NE),
and by DGESIC (Spain) under grant No. PB97-1261.
 
\end{titlepage} 
\addtocounter{page}{1} 
\paragraph{1.} 
The recent experimental measurements \cite{ktevna48} of direct CP 
violation in $K^0\to 2 \pi$ decays have led to a new world average  
\cite{ago:99} 
\begin{equation} 
\RE\,(\epoe) = (21.4\pm 4.0) \times 10^{-4} \,. 
\label{eq:exp} 
\end{equation} 
The theoretical status of $\epoe$ is reviewed in  
Refs.~\cite{buras:99,jamin:99}. Among the ingredients of the 
theoretical prediction of $\epoe$ within the standard model, we 
concentrate in this note on the quantity $\Omega_{IB}$ 
defined as 
\begin{equation}  
\Omega_{IB}:=\displaystyle\frac{\IM A_{2,IB}}{\omega \IM A_0}~. 
\label{eq:oib} 
\end{equation}  
We follow the conventional notation: 
the amplitudes $A_I (I=0,2)$ denote the $K \to \pi\pi$ amplitudes 
with isospin $I$ in the final state, the subscript $IB$ stands for 
isospin breaking due to $m_u\ne m_d$ (electromagnetic corrections are 
usually not included in $\Omega_{IB}$) and $\omega := \RE A_2/\RE A_0 
\approx 1/22.1$. In the theoretical analyses (e.g., in  
Refs.~\cite{buras:99,jamin:99}), one often takes 
\begin{equation}  
\Omega_{IB} \simeq \Omega_{\eta + \eta'} 
\end{equation} 
arising from \pe \, and $\eta$-$\eta'$ mixing \cite{etaetap}. 
 
In chiral perturbation theory (CHPT) \cite{wein:79,gl:84,gl:851},  
the effective field theory of the standard model at low energies,  
$\Omega_{IB}$ occurs already at lowest order, $O(p^2)$: 
\begin{equation}  
\Omega_{IB}= \displaystyle\frac{2 \sqrt{2} \epe^{(2)}}{3 \sqrt{3} 
\omega} = 0.13~, \label{eq:omegalo} 
\end{equation}  
where the lowest-order \pe \, mixing angle $\epe^{(2)}$ can be 
expressed in terms of quark mass ratios as 
\begin{equation}  
\epe^{(2)} = \displaystyle\frac{\sqrt{3}(m_d - m_u)}{4 (m_s - \hat m)} 
\label{eq:pe2} 
\end{equation}  
with $\hat m = (m_u + m_d)/2$ the average light quark mass. We use the 
canonical quark mass ratios \cite{hl:96} 
\begin{eqnarray}  
\displaystyle\frac{m_u}{m_d}= 0.55 \pm 0.04 ~,\qquad & & \qquad 
\displaystyle\frac{m_s}{m_d}= 18.9 \pm 0.8 ~. 
\label{eq:mqr} 
\end{eqnarray}  
 
The $\eta'$ contribution to $\Omega_{IB}$ is known to be large  
\cite{etaetap}. In the analysis of the Munich group 
\cite{buras:99,jamin:99,bosch:99}, a value $\Omega_{\eta + \eta'} 
=0.25 \pm 0.08$ is taken, about twice as big as the 
lowest-order value (\ref{eq:omegalo}). This raises the 
question about the size of possible other contributions competing  
with $\eta'$-exchange. It is the purpose of this letter to 
answer this question to $O(p^4)$ in CHPT. 
 
To show the sensitivity of $\epoe$ to 
$\Omega_{IB}$, we adopt an approximate formula of the Munich group 
\cite{buras:99,jamin:99,bosch:99} (not to be used for any 
``serious'' analysis, however) 
\begin{eqnarray} 
\frac{\ve'}{\ve} &\approx& 13\,\IM\lambda_t \biggl[ 
\frac{130\,\mbox{\rm MeV}}{m_s(m_c)}\biggr]^2  
\biggl[\, B_6^{(1/2)}(1-\Omega_{IB})  
 - 0.4 B_8^{(3/2)}\Big(\frac{m_t(m_t)}{165\, 
\mbox{\rm GeV}}\Big)^{2.5} 
\biggr]\biggl(\frac{\Lambda_{\overline{{\rm MS}}}^{(4)}}{340\, 
\mbox{\rm MeV}}\biggr) \,.\nn 
\label{eq:crude} 
\end{eqnarray} 
The so-called B-factors $B_6^{(1/2)}$, $B_8^{(3/2)}$ measure the 
deviation of the relevant hadronic matrix elements of four-quark 
operators from the vacuum saturation approximation. Since  
$\IM\lambda_t~B_6^{(1/2)} > 0$, a smaller value of $\Omega_{IB}$ 
implies a larger $\epoe$ and an increased sensitivity to the precise 
value of $\IM\lambda_t~B_6^{(1/2)}$. We come back to 
this formula at the end of this note.  
 
\paragraph{2.} 
To investigate \pe  \, mixing beyond leading order, we consider the 
inverse matrix propagator in the space of pseudoscalar octet 
fields ($\pi_3$, $\pi_8$): 
\begin{equation}  
\Delta(q^2)^{-1} = q^2 {\bf 1} - M_2^2 - \Pi(q^2) ~. 
\end{equation}   
The lowest-order mass matrix is given by 
\begin{equation}  
M_2^2 = B \biggl(\ba{cc} 2 \hat m & (m_u - m_d)/\sqrt{3} \\ 
(m_u - m_d)/\sqrt{3} & 2 (2 m_s + \hat m)/3 \ea \biggr) 
\end{equation}  
where $B$ is a low-energy constant of the lowest-order chiral 
Lagrangian related to the quark condensate \cite{gl:851}. 
The self-energy matrix $\Pi(q^2)$ of $O(p^4)$ has the simple form 
\begin{equation}  
\Pi(q^2) = C q^2 + D 
\end{equation}  
with symmetric matrices $C,D$ independent of the momentum $q$.  
 
The inverse matrix propagator can now be written as 
\begin{equation}  
\Delta (q^2)^{-1} = ({\bf 1} - C/2) \left[q^2 {\bf 1} - 
M_2^2 - D - \{C,M_2^2\}/2 \right] ({\bf 1} - C/2)~. 
\end{equation}  
We first diagonalize $M_2^2$, the mass matrix of $O(p^2)$, 
with an orthogonal matrix $O_2$ depending on the \pe \, mixing angle  
$\epe^{(2)}$ given in (\ref{eq:pe2}). Neglecting terms of higher order 
in $m_u - m_d$, we have 
\begin{eqnarray}  
O_2 &=& {\bf 1} + \epe^{(2)} \sigma \scs \sigma := i \sigma_2 \\ 
M_{2d}^2 &=& O_2 M_2^2 O_2^T = \mbox{\rm diag}(2B\hat m, 2 B (2 m_s + 
\hat m)/3)~.\no 
\end{eqnarray}  
 
It remains to diagonalize the resulting mass matrix in the basis of 
tree-level eigenstates, 
\begin{equation}  
M_{2d}^2 + O_2 D O_2^T + \{O_2 C O_2^T, M_{2d}^2\}/2 ~. 
\label{eq:mm}
\end{equation}  
This is achieved with another orthogonal matrix  
$O_4={\bf 1} + \epe^{(4)} \sigma$ with 
\begin{eqnarray}  
\epe^{(4)} &=& (M_\pi^2 - M_\eta^2)^{-1} \left[D_{38}+C_{38}(M_\pi^2+  
M_\eta^2)/2 +\epe^{(2)}(D_{88}-D_{33}) \right.\nn 
& & \left.+\epe^{(2)}(C_{88}-C_{33}) (M_\pi^2+M_\eta^2)/2 \right]~, 
\label{eq:pe4} 
\end{eqnarray}  
working as always up to $O(p^4)$ and neglecting terms of higher than 
first order in $m_u - m_d$. The expression in square brackets in  
(\ref{eq:pe4}) is just the off-diagonal 
element of the mass matrix (\ref{eq:mm}).
 
The inverse propagator now assumes its final form 
\begin{equation}  
\Delta (q^2)^{-1} = ({\bf 1} - C/2) O_2^T O_4^T  
\left(q^2 {\bf 1} - M_d^2\right) O_4 O_2 ({\bf 1} - C/2) 
\end{equation}  
with $M_d^2 = \mbox{\rm diag}(M_{\pi^0}^2,M_\eta^2)$. 
The transformation from the original fields ($\pi_3$, $\pi_8$) to the 
mass eigenfields ($\pi^0$, $\eta$) of $O(p^4)$ is therefore 
accomplished by a matrix 
\begin{eqnarray}  
V &=&  ({\bf 1} + C/2) O_2^T O_4^T  \nn 
&= & {\bf 1} - (\epe^{(2)} + \epe^{(4)})\sigma   
+ C/2 - \epe^{(2)} C \sigma /2 ~. 
\end{eqnarray}  
The matrix element of interest that contributes to all amplitudes 
involving \pe \, mixing to $O(p^4)$ is  
\begin{equation}  
V_{\pi_8 \pi^0} = \epe^{(2)}  + \epe^{(4)} + C_{38}/2 
+\epe^{(2)} C_{88}/2 ~. \label{eq:V80} 
\end{equation}  
 
Is the expression (\ref{eq:V80}) the generalization of the 
lowest-order mixing angle $\epe^{(2)}$ to $O(p^4)$? The answer is no 
because $(\ref{eq:V80})$ is in fact not a measurable quantity. Looking 
first at the last two terms,  
$C_{38}$ and $C_{88}$ are actually divergent. Moreover, whereas the  
$\eta$ mass shift $\Pi(M_\eta^2)= C_{88} M_\eta^2 + D_{88}$ is 
invariant under field redefinitions, $C_{88}$ is not. On the other 
hand, $\epe^{(4)}$ is a well-defined and therefore measurable 
quantity. It was first calculated in Ref.~\cite{gl:852} for the  
analysis of isospin violation in $K_{l3}$ form factors. Without 
electromagnetic corrections \cite{hh:95}, which are by definition not 
included in $\Omega_{IB}$, the result is 
\begin{eqnarray}  
\epe^{(4)} &=& \displaystyle\frac{2 \epe^{(2)}}{3(4\pi F_\pi)^2(M_\pi^2- 
M_\eta^2)} 
\biggl\{ 64(4\pi)^2 (M^2_K - M^2_\pi)^2 [3 L_7 + L^r_8(\mu)]\biggr. \nn 
&&  \mbox{} - M^2_\eta(M^2_K - M^2_\pi) \ln \frac{M^2_\eta}{\mu^2} 
- 2M^2_K(M^2_K - 2 M^2_\pi) \ln \frac{M_K^2}{\mu^2} \nn 
&& \biggl. \mbox{} + M_\pi^2 (M^2_K - 3 M^2_\pi) \ln \frac{M_\pi^2}{\mu^2} 
- 2M_K^2(M_K^2 - M_\pi^2) \biggr\}  
\label{eq:epe4} 
\eeqa 
where $F_\pi$ is the pion decay constant and  
$L_7, L_8^r(\mu)$ are low-energy constants of the chiral Lagrangian of  
$O(p^4)$ \cite{gl:851}. The scale dependence of $L_8^r(\mu)$ is of 
course cancelled by the chiral logarithms in (\ref{eq:epe4}). 
 
A possible definition of the \pe \, mixing angle up to $O(p^4)$ 
is therefore provided by 
\begin{equation}  
\epe := \epe^{(2)} + \epe^{(4)}~. 
\label{eq:epetot} 
\end{equation}  
In the notation of Ref.~\cite{gl:851}, the mixing angle 
(\ref{eq:epetot}) corresponds to $(\ve_1 + \ve_2)/2$. 
 
\paragraph{3.} 
After this general treatment of \pe \, mixing to $O(p^4)$, we now turn  
to the decays $K^0 \to \pi\pi$. As the previous discussion of the 
expression (\ref{eq:V80}) has shown, the non-measurable part 
$$ 
 C_{38}/2 +\epe^{(2)} C_{88}/2 
$$ 
must combine with other contributions of $O(p^4)$ specific to the 
decay $K^0 \to \pi^0 \pi^0$ to produce a measurable S-matrix 
element. This implies that there are additional contributions to the 
$K^0 \to \pi\pi$ decay amplitudes of $O[(m_u-m_d)p^2]$ that are not 
included in the \pe \, mixing angle (\ref{eq:epetot}). 
Some of these additional contributions have recently been considered 
in Ref.~\cite{gv:99}. The complete $K \to \pi\pi$ amplitudes to  
$O[(m_u-m_d)p^2]$ including electromagnetic corrections up to  
$O(e^2 p^2)$ \cite{cdg:9899} will be presented and analysed elsewhere 
\cite{emnp:00}. 
 
Here, we are concerned with the contribution of the \pe \, mixing 
angle to the quantity $\Omega_{IB}$. In fact, we can 
demonstrate that there are no other contributions of the type 
$(m_u-m_d)~L_i$ in $\Omega_{IB}$, where the $L_i$ are the ten 
low-energy constants in the chiral Lagrangian of $O(p^4)$ 
\cite{gl:851}. The first observation is that the 
strong chiral Lagrangian of $O(p^4)$ does not generate vertices with 
three mesons. Therefore, only the 
bilinear terms appearing in the two-point functions can contribute 
to the decays $K \to \pi \pi$ at $O(p^4)$. In addition to $\epe$,
$\Omega_{IB}$ contains the following combination of  
self-energy matrix elements: 
\begin{equation}  
C_{38} + \epe^{(2)} (C_{88} - C_{33})~. 
\label{eq:comb} 
\end{equation}  
The explicit dependence of these matrix elements on the 
constants $L_i$ is given by 
\begin{eqnarray}  
C_{38}(L_i) &=& - \displaystyle\frac{8 B}{\sqrt{3}F^2}~L_5 (m_u-m_d)~, \nn  
C_{33}(L_i) &=& - \displaystyle\frac{16 B}{F^2}\{L_4 (m_s+2\hat m)+L_5 
\hat m \}~, \nn 
C_{88}(L_i) &=&- \displaystyle\frac{16 B}{F^2}\{L_4 (m_s+2\hat m)+L_5 
(2m_s+\hat m)/3 \}~. 
\end{eqnarray} 
With the help of (\ref{eq:pe2}) one finds that the combination 
(\ref{eq:comb}) is indeed independent of the $L_i$. 
 
Therefore, the complete dependence of $\Omega_{IB}$ on the  
strong low-energy constants of $O(p^4)$ is contained in the \pe \, 
mixing angle (\ref{eq:epetot}) and we arrive at our final result 
\begin{equation}  
\Omega_{IB}^{\pi^0\eta}=\displaystyle\frac{2 \sqrt{2} \epe} 
{3 \sqrt{3} \omega}~. 
\end{equation} 
The superscript in $\Omega_{IB}^{\pi^0\eta}$ serves as a reminder  
that there are other isospin-violating contributions to 
$\Omega_{IB}$ in addition to \pe \, mixing. 
 
\paragraph{4.} 
For the numerical discussion, let us first look at the 
contributions of the low-energy constants $L_7$, $L_8$. As is 
well known, to $O(p^4)$ the effect of the $\eta'$ is completely  
contained in $L_7$ \cite{gl:851}. Taking the standard (mean) value  
$L_7=-0.4\times 10^{-3}$, the contribution of the $\eta'$ to the \pe 
\, mixing angle normalized to the lowest-order value is 
\begin{equation}  
\epe^{(4)}(L_7)/\epe^{(2)}= 1.10~. 
\end{equation}  
In agreement with earlier calculations \cite{etaetap}, $\eta'$ 
exchange more than doubles the lowest-order \pe \, mixing angle. The 
surprise comes from the second contribution due to $L_8^r(M_\rho)$ for 
which we take again the standard value $0.9\times 10^{-3}$: 
\begin{equation}  
\epe^{(4)}(L_8^r(M_\rho))/\epe^{(2)}= - 0.83~. 
\end{equation}  
The remaining (loop) contributions in (\ref{eq:epe4}) almost 
cancel for $\mu=M_\rho$. Altogether, we obtain the (scale-independent)  
result 
\begin{equation}  
\Omega_{IB}^{\pi^0\eta} = 0.16 
\label{eq:omres} 
\end{equation}  
to be compared with the lowest-order value $\Omega_{IB} = 0.13$ in  
(\ref{eq:omegalo}). 
 
Before estimating the theoretical error, we briefly 
discuss the physical origin of the $L_8$ contribution that nearly 
cancels the $\eta'$ contribution encoded in $L_7$. For this purpose we 
recall that the phenomenological values of the $L_i^r(M_\rho)$ can be 
well understood in terms of meson resonance exchange \cite{egpr:89}. 
In particular, $L_8^r(M_\rho)$ is only sensitive to the 
octet scalar resonances. In the case at hand, it is the $a_0(983)$ that 
couples both to $\eta\pi^0$ and to an isospin-violating tadpole 
proportional to $m_u-m_d$. Therefore, $a_0$-exchange contributes to  
\pe \, mixing via $L_8$ and this is the only 
low-lying meson resonance contribution. However, we hasten to 
emphasize that the result (\ref{eq:epe4}) is a strict consequence of 
QCD to $O(p^4)$ in the low-energy expansion and is independent of any 
specific interpretation of the numerical value of $L_8^r$. 
 
The theoretical uncertainty of the result (\ref{eq:omres}) for 
$\Omega_{IB}^{\pi^0\eta}$ is dominated by the uncertainties of the  
low-energy constants. The combination $3 L_7 + L_8^r(M_\rho)$ can be  
determined  from two observables \cite{gl:851}: the deviation from  
the Gell-Mann--Okubo mass formula, which is well under control, and a  
quantity $\Delta_M$ related to the $O(p^4)$ corrections for the ratio  
$M_K^2/M_\pi^2$. With a generous upper limit $|\Delta_M| \le 0.2$  
(compared to $|\Delta_M| \le 0.09$ in \cite{gl:851} and
$\Delta_M = 0.065 \pm 0.065$ in \cite{hl:96}) to allow also for
higher-order corrections, one finds 
\begin{equation}  
3 L_7 + L_8^r(M_\rho)= (-0.25 \pm 0.25) \times 10^{-3}~. 
\end{equation}   
In comparison, the errors of both the Gell-Mann--Okubo discrepancy and 
the quark mass ratios (\ref{eq:mqr}) 
entering $\epe^{(2)}$ can be neglected. The final result for the 
contribution of \pe \, mixing to $\Omega_{IB}$ is 
\begin{equation}  
\Omega_{IB}^{\pi^0\eta} = 0.16 \pm 0.03 ~. 
\label{eq:omfinal} 
\end{equation}  
An independent estimate can be obtained from the analysis of $K_{l3}$ 
form factors \cite{gl:852}. From the experimentally measured ratio of 
the $K^+ \pi^0$ to $K^0 \pi^-$ form factors at $q^2=0$ one can 
directly extract the \pe \, mixing angle $\epe$, leading to  
$\Omega_{IB}^{\pi^0\eta}=0.19 \pm 0.06 $. The two values are consistent  
with each other. Since the latter value is obtained under the 
assumption that electromagnetic corrections \cite{hh:95} can be  
neglected we consider (\ref{eq:omfinal}) as our final result. 
 
\paragraph{5.}  
The contribution of \pe \, mixing to $\Omega_{IB}$ does not include  
all isospin-violating corrections in $K^0 \to \pi\pi$ decays. 
As defined here to $O(p^4)$ in the low-energy expansion of QCD, 
it gives rise to 
\begin{equation}  
\Omega_{IB}^{\pi^0\eta} = 0.16 \pm 0.03 ~. 
\end{equation}  
This value is smaller than the one used previously where only  
$\eta$-$\eta'$ mixing was included at $O(p^4)$ \cite{etaetap}. To 
assess the impact on $\ve' / \ve$, we adopt the approximate formula 
(\ref{eq:crude}) with the central values for the B-factors taken by
the Munich group \cite{buras:99,jamin:99,bosch:99}, 
$B_6^{(1/2)}=1$, $B_8^{(3/2)}=0.8$. 
Lowering $\Omega_{IB}$ from 0.25 
to 0.16 corresponds to an increase of $\ve'/ \ve$ by $21\%$, bringing 
the theoretical prediction \cite{buras:99,jamin:99,bosch:99} closer to 
the experimental value (\ref{eq:exp}). In addition, a smaller value 
for $\Omega_{IB}$ increases the sensitivity of $\ve' / \ve$ to the 
B-factor $B_6^{(1/2)}$. 

  It has been shown recently \cite{PP:99} that final state interactions
induce a strong enhancement of the theoretical $\ve'/ \ve$ prediction,
correcting the $I=0$ and $I=2$ $K\to\pi\pi$ amplitudes with
the multiplicative dispersive factors $\Re_0 = 1.41\pm 0.06$ and 
$\Re_2 = 0.92\pm 0.02$, respectively.
The bag factors used before, which do not include final state interactions,
get then modified to $B_6^{(1/2)}|_{FSI}=1.4$ and $B_8^{(3/2)}|_{FSI}=0.7$.
The isospin-violating contribution $B_6^{(1/2)}\Omega_{IB}$ corresponds to 
two final pions with $I=2$ and, therefore, should be multiplied by $\Re_2$;
thus, $B_6^{(1/2)}\Omega_{IB}|_{FSI}= 0.15$.
The overall effect is to enhance $\ve'/ \ve$ by a factor 2.3.
The so-called ``central'' value in Refs.~\cite{buras:99,jamin:99,bosch:99},
$\ve'/ \ve = 7.0 \times 10^{-4}$, gets then increased to
$16 \times 10^{-4}$, in better agreement with the experimental measurement.
 
\vspace*{1cm} 
\noindent 
We thank Heinz Rupertsberger for useful discussions. 
 
\vspace{1cm} 
 


\begin{thebibliography}{99} 
\bibitem{ktevna48} 
A. Alavi-Harati et al. (KTeV-Coll.), Phys. Rev. Lett. 83 (1999) 22  
[\hepex{9905060}];\\ 
V. Fanti et al. (NA48-Coll.), \PL{B 465}{99}{335} [\hepex{9909022}]. 
\bibitem{ago:99} 
G. D'Agostini, \hepex{9910036}. 
\bibitem{buras:99} 
A.J.~Buras, \hepph{9908395}, Talk given at {\bf KAON 99}, 
Chicago, June 1999. 
\bibitem{jamin:99} 
M. Jamin, \hepph{9911390}, Talk given at {\bf 8th Int. Symposium on  
Heavy Flavour Physics}, Southampton, July 1999. 
\bibitem{etaetap} 
J.F. Donoghue, E. Golowich, B.R. Holstein and J. Trampetic, 
\PL{B 179}{86}{361};\\ 
A.J. Buras and J.-M. G{\'e}rard, \PL{B 192}{87}{156};\\ 
H.-Y. Cheng, \PL{B 201}{88}{155};\\ 
M. Lusignoli, \NP{B 325}{89}{33}. 
\bibitem{wein:79} 
S. Weinberg, Physica 96A (1979) 327. 
\bibitem{gl:84} 
J. Gasser and H. Leutwyler, Annals of Physics 158 (1984) 142. 
\bibitem{gl:851} 
J. Gasser and H. Leutwyler, \NP{B 250}{85}{465}. 
\bibitem{hl:96} 
H. Leutwyler, \PL{B 378}{96}{313}. 
\bibitem{bosch:99} 
S. Bosch et al., \hepph{9904408}, to appear in Nucl. Phys. B. 
\bibitem{gl:852} 
J. Gasser and H. Leutwyler, \NP{B 250}{85}{517}. 
\bibitem{hh:95} 
H. Neufeld and H. Rupertsberger, Z. Phys. C 68 (1995) 91;  
Z. Phys. C 71 (1996) 131. 
\bibitem{gv:99} 
S. Gardner and G. Valencia, \hepph{9909202}. 
\bibitem{cdg:9899} 
V. Cirigliano, J.F. Donoghue and E. Golowich, \PL{B 450}{99}{241} 
[\hepph{9810488}]; \hepph{9907341}; \hepph{9909473}. 
\bibitem{emnp:00} 
G. Ecker et al., in preparation. 
\bibitem{egpr:89} 
G. Ecker, J. Gasser, A. Pich and E. de Rafael, \NP{B 321}{89}{311}. 
\bibitem{PP:99} E. Pallante and A. Pich, \hepph{9911233}. 
 
\end{thebibliography}
\end{document}